\documentclass[twocolumn,english,prl,superscriptaddress,noshowpacs,amssymb]{revtex4}
\usepackage[T1]{fontenc}
\usepackage{amsmath}
\usepackage{amssymb}
\usepackage{graphicx}
\usepackage{pdfpages}

\makeatletter
\@ifundefined{textcolor}{}
{%
 \definecolor{BLACK}{gray}{0}
 \definecolor{WHITE}{gray}{1}
 \definecolor{RED}{rgb}{1,0,0}
 \definecolor{GREEN}{rgb}{0,1,0}
 \definecolor{BLUE}{rgb}{0,0,1}
 \definecolor{CYAN}{cmyk}{1,0,0,0}
 \definecolor{MAGENTA}{cmyk}{0,1,0,0}
 \definecolor{YELLOW}{cmyk}{0,0,1,0}
 }

\usepackage{bm}\@ifundefined{definecolor}{\usepackage{color}}{}

\usepackage[T1]{fontenc}
\usepackage[latin9]{inputenc}
\usepackage{textcomp}
\usepackage{color}

\usepackage{babel}

\makeatother

\begin{document}

\title{Superconducting quantum node for entanglement and storage of microwave radiation}

\author{E. Flurin}

\affiliation{Laboratoire Pierre Aigrain, Ecole Normale Sup\'erieure-PSL Research University, CNRS, Universit\'e Pierre et Marie Curie-Sorbonne Universit\'es, Universit\'e Paris Diderot-Sorbonne Paris Cit\'e, 24 rue Lhomond, 75231 Paris Cedex 05, France}

\author{N. Roch}

\affiliation{Laboratoire Pierre Aigrain, Ecole Normale Sup\'erieure-PSL Research University, CNRS, Universit\'e Pierre et Marie Curie-Sorbonne Universit\'es, Universit\'e Paris Diderot-Sorbonne Paris Cit\'e, 24 rue Lhomond, 75231 Paris Cedex 05, France}

\author{J.D. Pillet}

\affiliation{Laboratoire Pierre Aigrain, Ecole Normale Sup\'erieure-PSL Research University, CNRS, Universit\'e Pierre et Marie Curie-Sorbonne Universit\'es, Universit\'e Paris Diderot-Sorbonne Paris Cit\'e, 24 rue Lhomond, 75231 Paris Cedex 05, France}
\affiliation{Coll\`ege de France, 11 place Marcelin Berthelot, 75005 Paris, France}

\author{F. Mallet}

\affiliation{Laboratoire Pierre Aigrain, Ecole Normale Sup\'erieure-PSL Research University, CNRS, Universit\'e Pierre et Marie Curie-Sorbonne Universit\'es, Universit\'e Paris Diderot-Sorbonne Paris Cit\'e, 24 rue Lhomond, 75231 Paris Cedex 05, France}

\author{B. Huard}

\email[corresponding author: ]{benjamin.huard@ens.fr}

\selectlanguage{english}

\affiliation{Laboratoire Pierre Aigrain, Ecole Normale Sup\'erieure-PSL Research University, CNRS, Universit\'e Pierre et Marie Curie-Sorbonne Universit\'es, Universit\'e Paris Diderot-Sorbonne Paris Cit\'e, 24 rue Lhomond, 75231 Paris Cedex 05, France}

\date{\today}
\pacs{}

\begin{abstract} \textbf{
Superconducting circuits and microwave signals are good candidates to realize quantum networks, which are the backbone of quantum computers. We have realized a quantum node based on a 3D microwave superconducting cavity parametrically coupled to a transmission line by a Josephson ring modulator. We first demonstrate the time-controlled capture, storage and retrieval of an optimally shaped propagating microwave field, with an efficiency as high as 80~\%. We then demonstrate a second essential ability, which is the timed-controlled generation of an entangled state distributed between the node and a microwave channel.}\end{abstract}

\maketitle

Microwave signals are a promising resource for quantum information processing. Coupled to various quantum systems~\cite{Haroche2006,Schoelkopf2008,Xiang2013,Devoret2013a} they could realize quantum networks for continuous variable states, in which entangled information is processed by quantum nodes and distributed through photonic channels~\cite{Cirac1997,Kimble2008}. The quantum nodes should generate and distribute microwave entangled fields while controlling their emission and reception in time. Superconducting circuits are able to generate entanglement~\cite{Eichler2011,Wilson2104,Flurin2012a,Menzel2012a} and quantum memories provide control in time as demonstrated in emerging implementations in the microwave domain using spin ensembles~\cite{Wu2010,Kubo2012,Saito2013}, superconducting circuits~\cite{Yin2013,Wenner2013} or mechanical resonators~\cite{Palomaki2013,Palomaki2013a}. Here, we present a superconducting device both able to store and generate entangled microwave radiations shared between a memory and a propagating mode. It is based on the Josephson ring modulator~\cite{Bergeal,Bergeal2010a}  that enables to switch dynamically on or off the coupling between a low-loss cavity and a transmission line by frequency conversion. We demonstrate the time-controlled capture, storage and retrieval of a propagating coherent state in a long lived electromagnetic mode. Exploiting the versatility of this circuit, we then demonstrate the timed-controlled generation of an Einstein-Podolsky-Rosen (EPR) state distributed between the quantum memory and a propagating wave-packet. These new capabilities pave the way for complex quantum communication and quantum computing protocols by means of photonic channels in the microwave domain.

\begin{figure}[h!]
\includegraphics[scale=0.29]{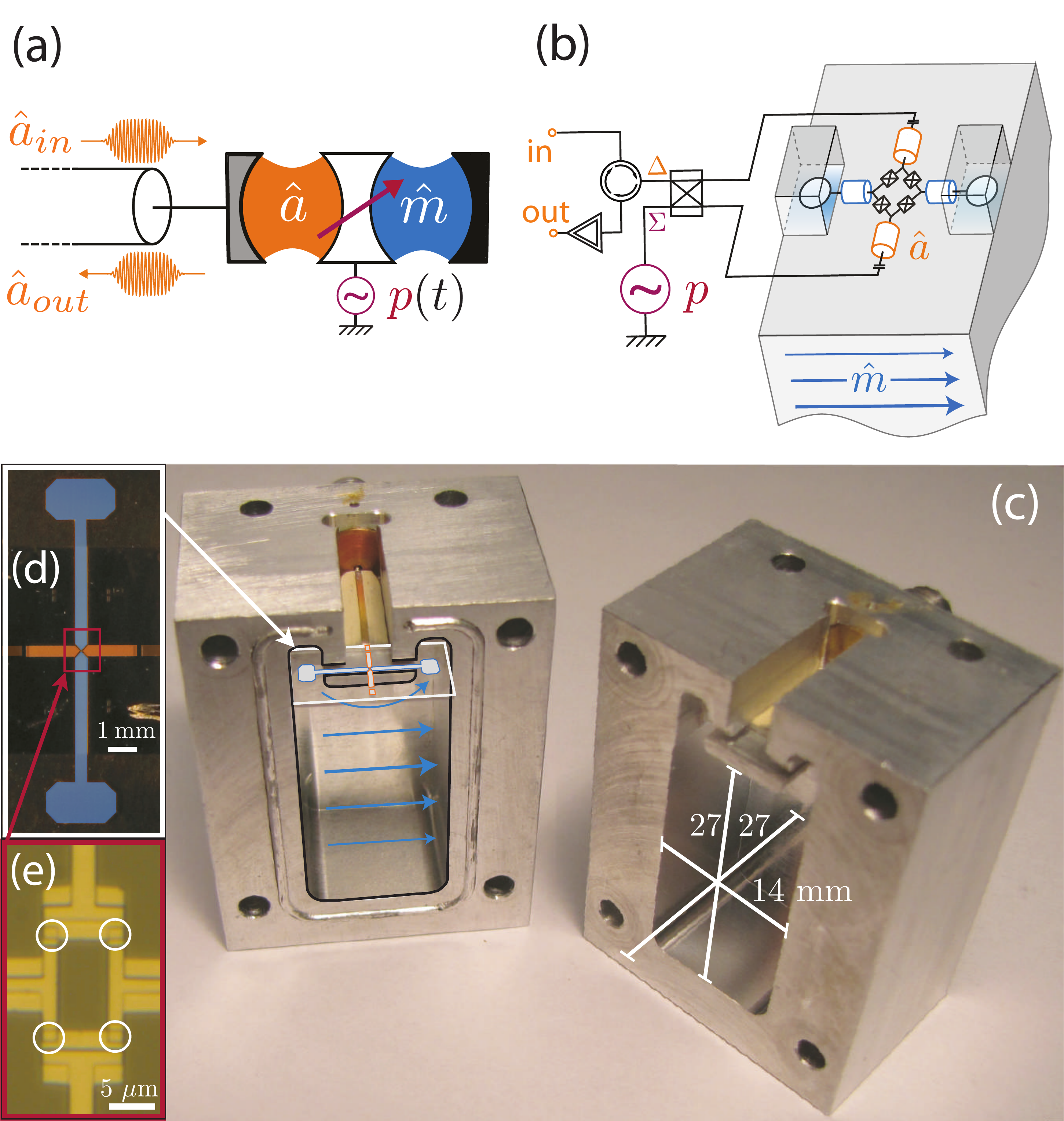}\caption{(a) Schematic of the experimental setup. A high-Q memory mode $\hat{m}$ is parametrically coupled to a low-Q buffer mode $\hat{a}$, hence to input/output propagating modes $\hat{a}_{in}$ and $\hat{a}_{out}$, depending on the pump amplitude $p$. (b),(c) Schematics and picture of the device. The on-chip circuit couples to a 3D superconducting cavity via antennas. The blue arrows represent the polarization of the fundamental mode TE 110 in the cavity. The Josephson ring and buffer resonator are on-chip. The differential mode ($\Delta$) couples with the buffer mode while the common mode ($\Sigma$) is used for addressing the pump. (d) Picture of the aluminum circuit fabricated on a c-plane sapphire substrate. The antennas (blue) and the buffer microstrip resonator (orange) are highlighted in false color. (e) Optical microscope image of the Josephson ring at the crossing between antennas and buffer resonator. The Josephson junctions are circled in white. \label{fig1}}
\end{figure}

The superconducting node is made of three components: a memory, a buffer and a parametric coupler linking them. The memory is the fundamental mode $\hat{m}$ at frequency $f_m=7.80~\mathrm{GHz}$ of a low-loss 3D superconducting cavity cooled down to $40\ \mathrm{mK}$ (Fig.~\ref{fig1}). The buffer is the fundamental mode $\hat{a}$ at frequency $f_a$ of an on-chip resonator and is the only component directly coupled to the network channels with propagating modes $\hat{a}_{in/out}$. The large coupling rate $\kappa_a=(20~\mathrm{ns})^{-1}$ between buffer and channel ensures fast communication compared to decoherence. The memory and buffer are parametrically coupled through a ring of four Josephson junctions pumped with a classical control field $p$ at frequency $f_p$. The magnetic flux through the ring allows to tune $f_a$ between 8.7 and 9.6~GHz. As described in previous works~\cite{Abdo2013}, the ring performs three-wave-mixing and $H_{mix}=\hbar \chi(\hat{a}+\hat{a}^{\dagger})(\hat{m}+\hat{m}^{\dagger})(p+p^{*})$. The device can be operated in two distinct ways depending on the pump frequency. For $f_p=|f_a-f_m|$, the device operates as a converter~\cite{Bergeal,Abdo2012}. In the rotating wave approximation (RWA) and with $p>0$ the term $H_{conv}=\hbar \chi p(\hat{a}^{\dagger}\hat{m}+\hat{a}\hat{m}^{\dagger})$ provides a tunable coupling rate $\chi p$  with frequency conversion between the buffer and memory modes. Conversely, for $f_p=f_a+f_m$, the RWA leads to the parametric down-conversion Hamiltonian $H_{pd}=\hbar \chi p(\hat{a}^{\dagger}\hat{m}^\dagger+\hat{a}\hat{m})$. The device then operates as an entanglement generator~\cite{Flurin2012a}. Starting from the vacuum state, an EPR state is distributed between the propagating mode $\hat{a}_{out}$ and memory mode $\hat{m}$.  These properties offer a striking resemblance with memories based on mechanical resonators. However, unlike the latter, superconducting circuits do not require extra cooling steps and offer three orders of magnitude larger input/output rates~\cite{Palomaki2013a,Palomaki2013}.

\begin{figure}
\includegraphics[scale=0.41]{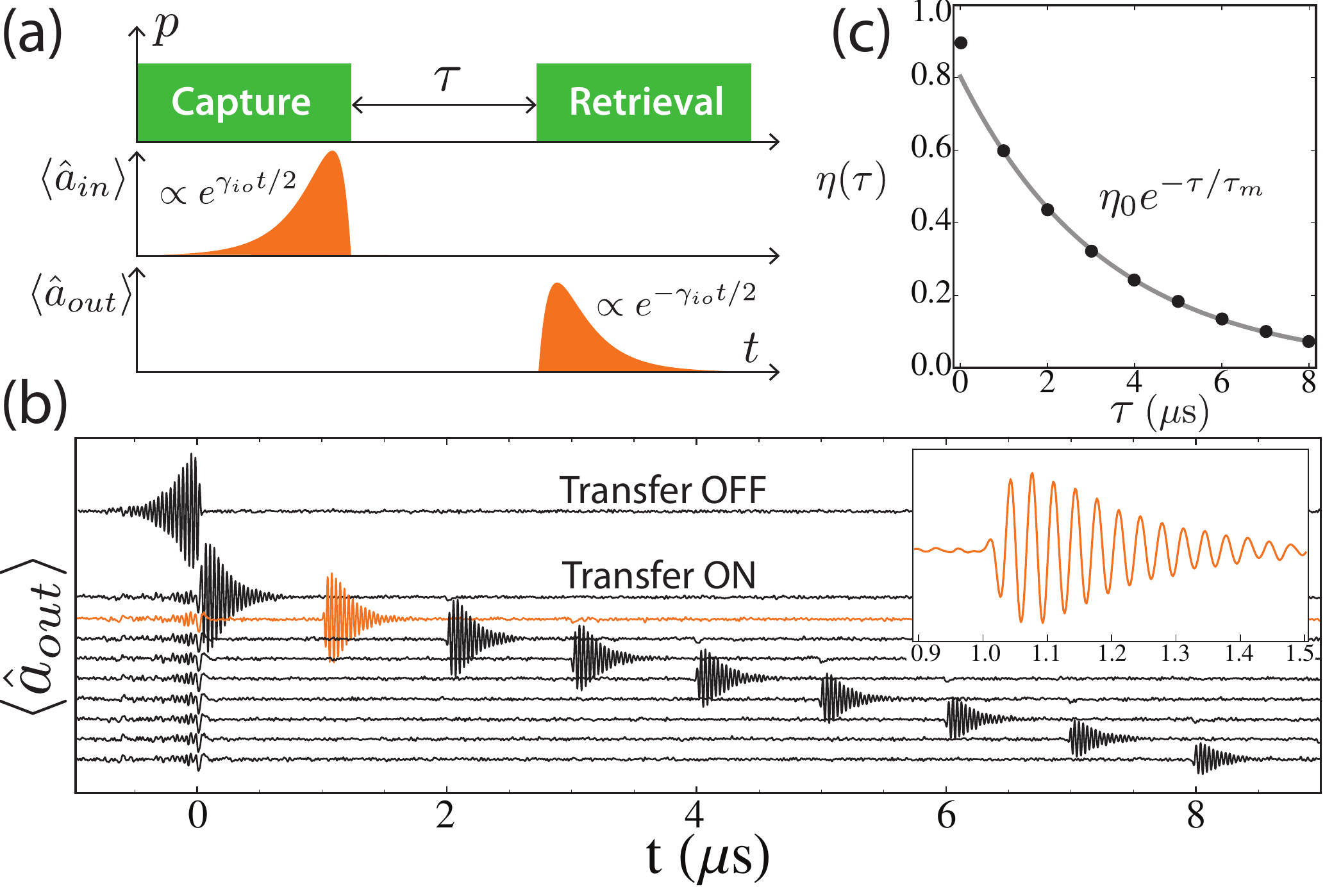}\caption{(a) Capture, store and release protocol. Pulse sequences for the pump field $p$ (green) at the difference frequency $f_p=f_a-f_m$, the input field $a_{in}$ and the resulting output field $a_{out}$ (orange). The temporal shape of the input field is chosen in order to optimize the capture efficiency. (b) Time traces of the amplitude of the output field down converted to $40 \mathrm{MHz}$ and averaged $6\times10^4$ times. The top trace is measured without pump and reveals the optimized input signal. The following traces correspond to the sequence of (a) with increasing delay $\tau$ between capture and retrieval from $0\ \mathrm{\mu s}$ to $8\ \mathrm{\mu s}$. The trace at $\tau=1~\mu\mathrm{s}$ is magnified for clarity as an inset. (c) Dots: retrieval efficiency $\eta$ as function of delay $\tau$. $\eta$ is defined as the ratio of the retrieved energy normalized to the input energy. Plain line: exponential decay $\eta_0 e^{-\tau/\tau_{m}}$ characterizing the memory lifetime. Best fit obtained for $\eta_0=80~\%$ and $\tau_m=3.3~\mu\mathrm{s}$.\label{fig2}}
\end{figure}

In order to demonstrate the performances of the memory, one can first capture and retrieve a propagating classical field. Depending on its temporal shape, the pump amplitude has to be shaped appropriately in order to maximize the capture efficiency, while taking care of the resonance frequency shifts that are induced by the pump power~\cite{Novikova2007,Sete2014,supmat}. In this first experiment, we used the easier dual approach of optimizing the temporal shape of  an incoming coherent state so that it is captured by a square pump pulse turning off at time $t=0$ (Fig.~\ref{fig2}a) similarly to Ref.~\cite{Yin2013,Wenner2013}. The optimal shape corresponds to the time-reverse of a signal retrieved from an initially occupied memory~\cite{supmat, Wenner2013}. It is an exponential rise in power (Fig.~\ref{fig2}), which defines the input/output rate $\gamma_{io}$ of the quantum node. Note that, due to the presence of the buffer, the pulse finally turns off smoothly on the scale of $\kappa_a^{-1}$ \cite{supmat}. The amplitude of the pump pulse was chosen in order to maximize the input/output rate to $\gamma_{io}$. Indeed, for large enough pump powers such that $\chi p>\kappa_a/4$ the modes $\hat{a}$ and $\hat{m}$ hybridize and the input/output rate saturates to its upper limit~\cite{supmat}. In practice, the finite coupling capacitance between the 3D cavity and the antennas (Fig.~\ref{fig1}) prevents $\gamma_{io}$ to reach an upper limit of $\kappa_a/2$ and we found a three times lower rate $\gamma_{io}\approx(110\ \mathrm{ns})^{-1}$ instead \cite{supmat}. It is worthwhile to note that, although the memory has a finite lifetime, the frequency conversion between modes $m$ and $a$ ensures that the input/output rate is exactly zero $\gamma_{io}=0$ when the pump is turned off, leading to an infinite on/off ratio. Besides, the device being non-resonant with the conversion operating frequency $f_p=f_a-f_m\approx 1.5\ \mathrm{GHz}$, the transfer rate can be varied much faster than $\kappa_a$.

The amplitude $\langle \hat{a}_{out}\rangle$ of the mode coming back from the device is measured for several pump pulse sequences (Fig.~\ref{fig2}b). In a first control measurement (top trace), the pump is kept turned off such that the measurement corresponds to the directly reflected incoming pulse~\footnote{Note that the signal is 20~MHz out of resonance with the buffer mode when the pump is turned off, so that the reflected power is identical to the incoming power.}. In the following measurements (traces below) the pump is turned on before time $0$ and after time $\tau$ (Fig.~\ref{fig2}a). A small part of the incoming pulse energy is reflected, at least 5~\% according to the average trace in Fig.~\ref{fig2}a, while it is sent at $t<0$ indicating the efficient absorption of this pulse shape. When the pump is turned back on after a delay $\tau$, the device releases the captured state back in the transmission line as can be seen in Fig.~\ref{fig2}b. Note that the chosen temporal shape of the incoming signal is indeed the time reverse of these pulses up to an amplitude rescaling, which corresponds to the efficiency of the memory. Calculating the memory efficiency $\eta$, which is the ratio between the retrieved pulse energy and the incoming pulse energy leads to an exponential decay as a function of delay time $\eta(\tau)=\eta_0 e^{-\tau/\tau_{m}}$ (Fig.~\ref{fig2}c). The memory lifetime $\tau_m=3.3~\mu\mathrm{s}$ is much larger than $\gamma_{io}^{-1}$ but limited by unidentified losses in the 3D cavity coupled to the antennas. The much smaller decay rates achieved in similar 3D cavities~\cite{Paik2011a} leave room for improvement in the future. Note that the anomalously large efficiency at zero delay $\eta(0)>\eta_0=80~\%$ should not be considered as a useful efficiency. Indeed, part of the retrieved energy right after time $t=0$ corresponds in fact to the decay of the part of the signal that was not stored in the memory mode and stayed in the buffer mode instead. Note that the memory withstands large signal amplitudes, since there are here 10 photons on average in the incoming wavepacket. Besides the outgoing phase is identical to that of the incoming pulse. The phase coherence properties of the device at the single photon level are demonstrated below. Finally, the number of operations that can be performed by the memory within its lifetime is limited by the time-bandwidth product $\gamma_{io}\tau_m= 30$. This combination of large memory efficiency and time-bandwidth product makes this device a state of the art quantum memory~\cite{Simon2010,Wenner2013}.

Promisingly, the device cannot only be used as a memory but also as an entanglement generator. In a second experiment, we demonstrate the generation of an EPR state distributed between the propagating mode $\hat{a}_{out}$ and the memory mode. Note that this experiment has been performed during another cooldown of the same device for which the memory lifetime was slightly degraded to $\tau_m=2.3~\mu\mathrm{s}$. Starting from the vacuum state both in the memory and in the mode $\hat{a}_{out}$, a pulse at pump frequency  $f_p=f_a+f_m= 17.28~\mathrm{GHz}$ produces a two-mode squeezed vacuum state $\left|Sq\right\rangle =e^{iH_{pd}\tau/\hbar}|0\rangle_{a}|0\rangle_m=\cosh(r)^{-1}\sum\tanh(r)^n\left|n\right\rangle _{a}\left|n\right\rangle _{m}$ where the squeezing parameter $r$ increases with the pump pulse amplitude~\cite{Flurin2012a}. The entanglement between memory and propagating modes can be demonstrated by measuring the correlations between the fluctuations of their mode quadratures, and showing that there is more correlation than allowed by classical physics~\cite{Eichler2011,Wilson2104,Menzel2012a}. The quadratures of both modes can be measured using the same detector on line $a$ provided that the memorized field is released into the transmission line at a later time.

\begin{figure}
\includegraphics[scale=0.4]{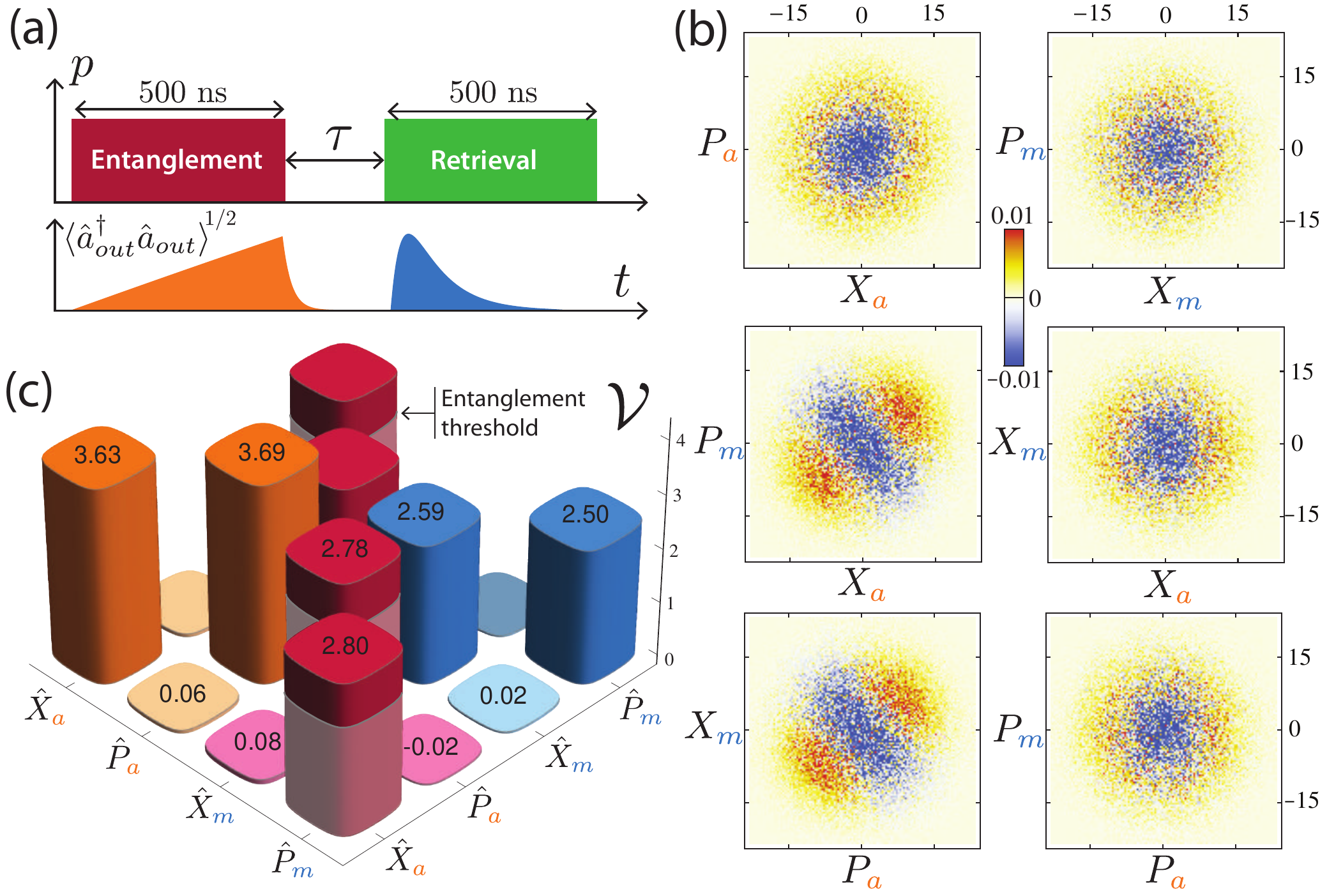}\caption{Entanglement between memory and propagating mode. (a) Scheme of the pulse sequence. Top: pump amplitude $p$ is shown in red for $f_p=f_a+f_m$ and in green for $f_p=f_a-f_m$. Bottom: output noise amplitude in time. (b) Difference between measured quadrature histograms with pump turned on and off for all pairs of quadratures for a delay $\tau=200~\mathrm{ns}$. The quadratures are calibrated separately (see text) so that the vacuum fluctuations would result in $\langle X_{a,m}^2\rangle=\langle P_{a,m}^2\rangle=1/4$. The color scale is in units of the maximal value of the histogram with pump off. (c) Resulting two-mode covariance matrix. The convention used is such that the vacuum state corresponds to the unity matrix. The $2\times2$ block-diagonal matrices in orange and blue represent the single mode $\hat{a}$ and $\hat{m}$ covariance matrices. The off-diagonal matrices in red represent the correlations between modes. The typical error on the matrix terms is 0.07. Correlations go beyond the greyed regions which demonstrates entanglement. \label{fig3}}
\end{figure}

The pulse sequence used in the experiment (Fig.~\ref{fig3}a) starts by a square pump pulse at $f_p=f_a+f_m=17.28~\mathrm{GHz}$ during $500\ \mathrm{ns}$ that generates an EPR state. While one part of the pair is stored in the memory, the other part propagates in the transmission line, is amplified by a low-noise amplifying detection setup and recorded using fast digital heterodyne detection based on a Field Programmable Gate Array (FPGA)~\cite{Eichler2011,Eichler2012}. After a delay $\tau=200\ \mathrm{ns}$, a square pulse is applied on the pump field at $f_p=f_a-f_m$ with an amplitude such that the output rate is $\gamma_{io}$ and lasting for $500\ \mathrm{ns}$. This pulse releases the memory field which is then amplified and measured using the heterodyne detection setup. At the end of a sequence, the four mode quadratures $\hat{X}_a$, $\hat{P}_a$, $\hat{X}_m$ and $\hat{P}_m$ have been measured (defining $\hat{X}_m\equiv(\hat{m}+\hat{m}^\dagger)/2$ and $\hat{P}_m\equiv(\hat{m}-\hat{m}^\dagger)/2i$). 

In order to perform a tomography of the entangled state, this sequence is repeated $4\times10^7$ times. The FPGA generates six histograms giving the probability distribution of measurement outcomes as a function of every pair of mode quadratures (see phase space in Fig.~\ref{fig3}b). The total acquisition and processing time amounts to 5 minutes. In practice, the resulting histograms are dominated by the uncorrelated noise of the detection setup. It is possible to cancel the contribution of this noise background by turning off the pump in another pulse sequence and subtracting the corresponding histograms to the original ones (Fig.~\ref{fig3}b)~\cite{Eichler2011,Eichler2012}. In order to avoid gain drifts, on and off pump sequences are performed every 30~s.

In the single-mode histograms along $(\hat{X}_a,\hat{P}_a)$ and along $(\hat{X}_m,\hat{P}_m)$, a phase-independent increase in the fluctuations is observed. This corresponds to the thermal state produced in a mode by tracing out the EPR state $|S_q\rangle$ on the other mode. On the contrary, the histograms along the quadratures of two different modes $(\hat{X}_a,\hat{P}_m)$ and $(\hat{P}_a,\hat{X}_m)$ exhibit large values along a diagonal indicating strong correlations between modes. For instance, the measurement outcomes $X_a$ and $P_m$ are likely to have the same sign. These correlations need to be characterized quantitatively in order to demonstrate whether or not they are non-classical.

\begin{figure}
\includegraphics[scale=0.40]{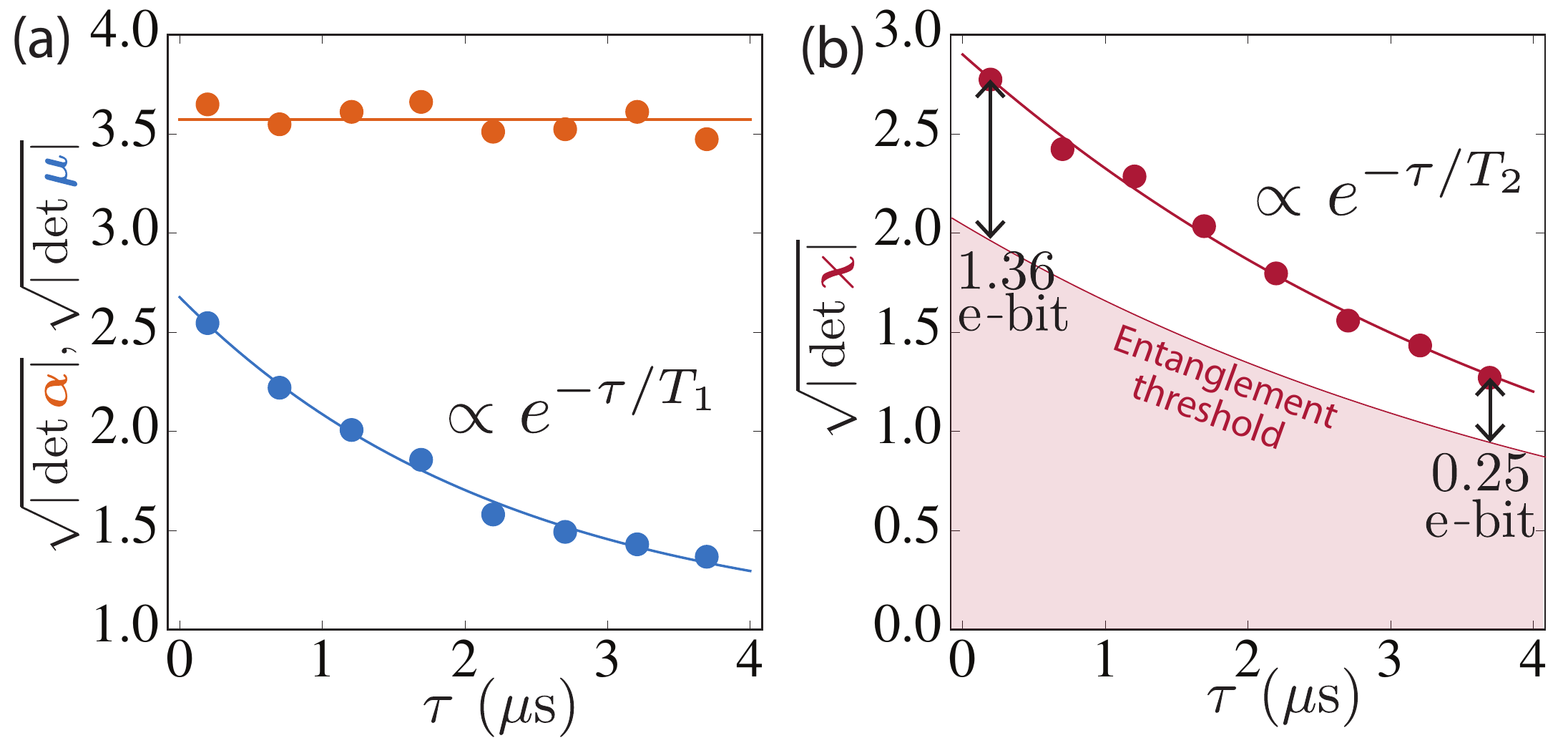}\caption{Covariance matrix and entanglement as a function of the storage time $\tau$. (a) Dots: diagonal terms of the covariance matrix $\mathcal{V}$ giving the energy of each mode. Lines: average value (for $\alpha$) and exponential fit (for $\beta$). The decay rate of the terms in $\beta$ gives the energy relaxation time $T_1=2.3\pm0.1\ \mathrm{\mu s}$. (b) Dots: Off-diagonal amplitudes in $\mathcal{V}$ representing the coherence between memory and propagating modes. Line: exponential fit, whose rate sets the decoherence time $T_2=4.5\pm0.1\ \mathrm{\mu s}$. Correlations above the entanglement threshold demonstrate entanglement between memory and propagating modes. The threshold decays in time because of the single mode noise decay in (a). Logarithmic negativity $E_N$ measuring the entanglement between modes is indicated for the two limit times.  \label{fig4}}
\end{figure}

The correlations can be calibrated using the known variance of the single mode quadratures. Indeed, for mode $a$, the thermal state corresponds to amplified vacuum fluctuations with a power gain $\cosh(2r)$ resulting in a variance for both quadratures $\Delta X_a^2=\Delta P_a^2=\cosh(2r)/4$. Note that we assume that the field is in the vacuum at thermal equilibrium with the refrigerator temperature $45~\mathrm{mK}\ll hf/k_B\approx 0.4~\mathrm{K}$~\footnote{We have consistently found less than 0.5~\% of excitation in 3D modes with similar setups and cavities using a 3D transmon qubit as a photocounter.}. The calibration then comes down to determining the gain $\cosh(2r)$ precisely. This can be done by storing a small coherent field (about 1 photon on average) in the memory and measuring the output amplitudes with and without applying the entangling $500~\mathrm{ns}$ pump tone at $f_p=17.28~\mathrm{GHz}$ before release~\cite{Flurin2012a}. The entangling pulse effectively amplifies the coherent field with an amplitude gain $\cosh(r)$ which is here found to be equal to $1.51$.

One can then calculate the covariance matrix $\mathcal{V}$ of the two mode state (Fig.~\ref{fig3}c), which fully characterizes the EPR state since it is Gaussian with zero mean~\cite{Braunstein2005}. The FPGA processes $4\times10^7$ pulse sequences in 5 minutes so that $\mathcal{V}$ is calculated with minimal post-processing~\cite{Eichler2011,Adesso2005a,Menzel2012a,supmat}. In a coordinate system where $\textbf{x} =\{\hat{X}_a,\hat{P}_a,\hat{X}_m, \hat{P}_m\}$, one defines $\mathcal{V}_{ij}=2(\langle x_i x_j +x_jx_i\rangle -2\langle x_i\rangle\langle x_j\rangle)$. Physically, it is meaningful to decompose it in four $2\times2$ block matrices.
\begin{equation}
\mathcal{V}=\left(
  \begin{array}{ c c }
    \boldsymbol{\alpha} & \boldsymbol{\chi} \\
     \boldsymbol{\chi}^T & \boldsymbol{\mu}
  \end{array} \right).
\end{equation}
The diagonal blocks $\boldsymbol\alpha$ and $\boldsymbol\mu$ are the single-mode covariance matrices for $\hat{a}$ and $\hat{m}$ respectively. Since an EPR state is thermal when disregarding the other mode, there is no correlation between quadratures $X$ and $P$ for a single mode and the variances $\Delta X^2$ and $\Delta P^2$ are almost equal. For mode $a$, by definition of the calibration process, one gets $\mathcal{V}_{11}\approx \mathcal{V}_{22}\approx\cosh(2r)=3.66$ (Fig.~\ref{fig3}c). The memory mode is less occupied because of losses at a rate $\tau_m^{-1}$ during the entanglement pulse and the waiting time $\tau=200~\mathrm{ns}$ so that $\mathcal{V}_{33}\approx \mathcal{V}_{44}\approx 2.55$. Conversely, the off-diagonal blocks $\boldsymbol\chi$ correspond to the correlations between modes. In each block, the phase of the pump field was optimized to put all the weight of the correlations in the two terms $\mathcal{V}_{14}\approx \mathcal{V}_{23}\approx 2.79$. The amount of entanglement in the two mode state can be measured by the logarithmic negativity $E_N$. It corresponds to an upper bound for distillable entanglement~\cite{Adesso2005a}. Here, the memory and the propagating modes share $E_N=1.36$ entangled bits (e-bits), which indeed demonstrates the ability of the device to generate and preserve entanglement between modes.

The experiment was repeated for various storage times $\tau$ (Fig.~\ref{fig3}a). The typical amplitude $\sqrt{|\mathrm{det}\boldsymbol\mu|}$ of the memory mode terms in $\mathcal{V}$ decrease exponentially with $\tau$ (Fig.~\ref{fig4}a) as expected from the experiment with coherent states in Fig.~\ref{fig2}c. This leads to a relaxation time for the memory of $T_1=2.3\pm0.1\ \mu\mathrm{s}$ in agreement with the memory lifetime $\tau_m$ measured using coherent states in the same cool down of the device. The small variations in the amplitude of the propagating mode $\sqrt{|\mathrm{det}\boldsymbol\alpha|}$ with $\tau$ give a sense of the measurement uncertainty (Fig.~\ref{fig4}a). Interestingly, the two-mode correlations also decay exponentially (Fig.~\ref{fig4}b). The corresponding characteristic time is the decoherence time $T_2=4.5\pm0.1\ \mu\mathrm{s}$ of the memory. The fact that $T_2\approx2 T_1$ demonstrates that energy relaxation dominates all decoherence mechanisms during the storage of a quantum state. The logarithmic negativity also decreases with $\tau$ as shown in Fig.~\ref{fig4}b.

In conclusion, we have realized quantum node based on an hybrid 2D/3D superconducting circuit. The efficient capture, storage and retrieval of a coherent state was demonstrated. Moreover, the device permits the generation and storage of entangled states distributed between the node and photonics channels. The versatility of the device paves the way for complex quantum communication protocols in the microwave domain such as continuous variable quantum teleportation. Besides, it provides a useful resource for 3D cavities where the on-demand extraction of a field quantum state was needed. This could be used to implement readout and feedback in cavity networks or even quantum computation with the memory field itself~\cite{Leghtas2013}. Finally, superconducting qubits can easily be embedded in this device, therefore enabling the strong coupling between a qubit and a continuous variable entangled state, which leads to protected quantum memories~\cite{Devoret2013a} and even protected quantum computing with microwave fields~\cite{Leghtas2013,Mirrahimi2013}.

\begin{acknowledgements}
We thank Michel Devoret, Vladimir Manucharyan, Mazyar Mirrahimi and Pierre Rouchon for enlightening discussions and Landry Bretheau and Philippe Campagne-Ibarcq for proofreading. Nanofabrication has been made
within the consortium Salle Blanche Paris Centre. This work was supported
by the EMERGENCES program from Ville de Paris under the project QUMOTEL and by the Idex ANR-10-IDEX-0001-02 PSL *. JDP acknowledges financial support from Michel Devoret. \end{acknowledgements}


\end{document}